\documentclass[
aps,
 ,twocolumn
]{revtex4}
\usepackage{graphicx}

\begin{document}

\title{Quantum Key Distribution without sending a Quantum Signal}

\author{T.C. Ralph and N. Walk}
\affiliation{
School of Mathematics and Physics, University
of Queensland, Brisbane, Queensland 4072, Australia}

\date{\today}

%

%

\begin{abstract}
{Quantum Key Distribution is a quantum communication technique in which random numbers are encoded on quantum systems, usually photons, and sent from one party, Alice, to another, Bob. Using the data sent via the quantum signals, supplemented by classical communication, it is possible for Alice and Bob to share an unconditionally secure secret key. This is not possible if only classical signals are sent. Whilst this last statement is a long standing result from quantum information theory it turns out only to be true in a non-relativistic setting. If relativistic quantum field theory is considered we show it is possible to distribute an unconditionally secure secret key without sending a quantum signal, instead harnessing the intrinsic entanglement between different regions of space time.  The protocol is practical in free space given horizon technology and might be testable in principle in the near term using microwave technology.}
\end{abstract}

\maketitle

\vspace{10 mm}

\section{Introduction}

When information is carried by quantum systems then the abstract rules of information science developed for classical systems are modified \cite{NIE00}. If we now consider information carried on relativistic quantum systems then the rules are modified further \cite{PER04,DOW12}. One aim of this new field of relativistic quantum information is to understand whether new, more powerful information protocols emerge in this new setting. Here we describe a specific protocol of this type. Whilst quantum communication protocols that rely on the impossibility of superluminal signalling have been described recently \cite{KEN11} this is the first time a fundamentally different protocol has been described that relies on the full machinery of relativistic quantum field theory.

The key physics we will rely on here are quantum correlations of space time which arise in relativistic quantum field theory. It is well known that the restriction of quantum field modes to the left and right Rindler wedges leads observers of these modes to find that the ground state as defined in terms of the normal Minkowski modes, appears as an entangled state between the two wedges \cite{UNR76, DAV75}. Physically, the Rindler modes couple to detectors travelling on uniformly accelerating trajectories. Alternatively one can restrict the modes to the future and past light cones. For the massless scalar fields we will consider here, observers again couple to the Rindler modes \cite{OLS11,OLS12}. Physically, in this case, the modes couple to time-like separated inertial detectors with time dependent energy level separations such that their ``clocks" follow conformal time. Coupling to the Rindler modes by inertial detectors that are space-like separated has also been studied \cite{reznik:2003}. 
Extraction of entanglement from the Rindler modes has  traditionally been studied in terms of Unruh-DeWitt detectors \cite{UNR76}, single two-level systems coupled isotropically to the field and placed into: accelerated motion \cite{UNR76}; switched rapidly \cite{reznik:2003}; or given a time dependent resonant frequency \cite{OLS11}. More recently coupling to single harmonic oscillators has been considered \cite{LIN06}. The extraction of entanglement from the vacuum field has been referred to as havesting of entanglement \cite{BRO13}. However, the rapid decay of the entanglement with distance \cite{reznik:2003} means that these techniques are not directly useful for quantum communication protocols.

In this paper we will describe how entanglement of the electromagnetic vacuum can be observed using inertial macroscopic field detectors, i.e. homodyne detection. The particular novelty of this approach comes from the directionality and efficiency of homodyne detection which allows highly pure entanglement to be detected over far greater distances. Even though the vacuum entanglement is not transferred to an independent quantum system it is still useful for quantum information protocols. In particular we describe an entanglement-based quantum key distribution protocol in which no quantum signals are exchanged by the participants and the security arises purely from the intrinsic entanglement of space-time.

\section{Rindler and Light Cone Coordinates}

We consider a $(3 + 1) D$ massless scalar field description. Minkowski co-ordinates, $(x_1, x_2, x_3, t)$, are the standard ones for describing inertial observers. Rindler co-ordinates, $(\xi, x_2, x_3, \eta)$, describe trajectories that are restricted either to the left or right Rindler wedges. The Minkowski and Rindler coordinate systems are related via \cite{BIR82}:
\begin{eqnarray}
t &= & {{1}\over{a}} e^{a \xi} \sinh(a \eta); \;\;\;\;\;\;\; x_1 = \pm {{1}\over{a}}  e^{a \xi} \cosh(a \eta); 
\label{R1}
\end{eqnarray}
where $+$ ($-$) corresponds to the right (left) wedge. A stationary observer in Rindler co-ordinates, sufficiently well localized around $\xi = 0$, follows a uniformly accelerated trajectory in Minkowski co-ordinates. The rate of acceleration is given by the parameter $a$. Throughout this paper we work in units for which $c=1$. An important feature of the Rindler transformations is that the fields continue to satisfy the wave equation in the new coordinates.

Alternatively we can consider transforming into modes restricted to the past and future light cones. The future-past co-ordinates, $(\epsilon, x_2, x_3, \tau)$, describe trajectories that are restricted either to the future or past light cones. The Minkowski and futute-past (F-P) coordinate systems are related via \cite{OLS11}:
\begin{eqnarray}
t &= & \pm {{1}\over{a}} e^{a \tau} \cosh(a \epsilon); \;\;\;\;\;\;\; x_1 = \pm {{1}\over{a}}  e^{a \tau} \sinh(a \epsilon); 
\label{R2}
\end{eqnarray}
where $+$ ($-$) corresponds to the future (past) light-cone. A stationary observer in light-cone co-ordinates, sufficiently well localized around $\epsilon = 0$, is stationary in Minkowski co-ordinates but follows a conformal time $\tau$ with $a$ parametrizing the difference between conformal and Minkowski time. A key observation from Ref.\cite{OLS11} is that detectors described by coordinate transformations of the form Eq.\ref{R1} {\it or} Eq.\ref{R2} couple to the {\it same} Rindler modes.

\section{Homodyne Detection of Rindler Entanglement}

We wish to consider two localized observers, Alice and Bob, whose detectors are stationary at the origin in the light cone coordinates defined by Eqs \ref{R2}, with $+$ ($-$) corresponding to Alice (Bob), who detect the quadrature amplitudes of the Minkowski vacuum. We use here the approach of Ref. \cite{DOW13} (note also the alternate approach described in Refs \cite{BRU10} and \cite{DRA12}). They perform homodyne detection on the vacuum, as seen in their reference frames, each using, as a local oscillator mode, a state with coherent amplitude $\beta$, where $\beta$ is real and $\beta >> 1$. Each homodyne detector is formed from two identical photo-detectors, that detect distinct modes 
after they have been mixed on a beamsplitter. The photocurrents from the photo-detectors are subtracted to give the output signal. The purpose of the local oscillator is to provide a shared classical phase reference, and we can regard the local oscillator beams as being independently generated by Alice and Bob using a combination of shared precision clocks and classical communication. 

To be consistent with the coordinates of Eqs \ref{R2}, all these optical elements must obey the Schr\"odinger equation \cite{OLS11}:
\begin{eqnarray}
i {{d \Phi}\over{d \tau}} = H_o \Phi \;\;\; \to \;\;\; i {{d \Phi}\over{d t}} = \pm {{H_o}\over{a t}} \Phi,
\label{SE}
\end{eqnarray}
where the right-hand side shows the equivalent of the conformal time expression in Minkowski time, with $+$ ($-$) for the future (past) coordinates of Alice (Bob) (see the Appendix for more details). The output of such a homodyne detector at some conformal time, $\tau$, is represented by the following operator \cite{BAC03}:
\begin{equation}
\hat{O}_J(\tau, \phi) = \hat b_{J, S}(\tau) \hat b_{J, L}^{\dagger}(\tau) e^{i \phi} +\hat b_{J, S}^{\dagger}(\tau)\hat b_{J, L} (\tau)e^{-i \phi}
\label{OR}
\end{equation}
where $\hat b_{J, K}$ ($\hat b_{J, K}^{\dagger}$) are boson annihilation (creation) field operators with $J=F,P$ for Rindler modes in the future or past light cones respectively. The subscripts $K=S,L$ refers to the signal and local-oscillator modes respectively. The relative phase $\phi$ determines the quadrature angle detected (see the Appendix for more details). 
Creation of a coherent state can be modelled as a unitary displacement of the vacuum. Physically, this state is an excellent approximation to the state produced by a well-stabilized laser. Once each coherent state has been independently, locally created by displacing the Minkowski vacuum, it is then ``chirped" (i.e. a rapid scanning of the instantaneous frequency) according to the scaling in Eq.\ref{SE} such that it effectively beats with the Future or Past modes. 
Alice's (Bob's) displacement operator can be written: $\hat D_F(\beta)=\exp[\beta(b_{F, D}^{\dagger}-b_{F, D})]$ ($\hat D_P(\beta)=\exp[\beta(b_{P, D}^{\dagger}-b_{P, D})]$) where the subscript $D$ labels the mode to which the displacement is perfectly matched. 

Generically, the mode operators can be spectrally decomposed as:
\begin{eqnarray}
\hat{b}_J = \int dk_d f_J(k_d) e^{-i \Omega_d \tau} \hat{b}_{J,k_d} \label{rinn}
\end{eqnarray}
where $k_d=(k_{d1},k_{d2},k_{d3})$ refers to Alice or Bob's detector wave-vector with the Rindler frequency given by $\Omega_d=\sqrt{k_{d1}^2+k_{d2}^2+k_{d3}^2}$. The integral $\int dk_d$ is over the whole wave-vector space. The operators $\hat{b}_{k_d}$ are single frequency Rindler operators, obeying the boson commutation relation
\begin{equation}
[\hat{b}_{k_d}, \hat{b}_{k'_d}^{\dagger}] = \delta(\Omega_{d}-\Omega'_{d})\delta(k_{d2}-k'_{d2})\delta(k_{d3}-k'_{d3}).
\label{Rcomm}
\end{equation}
%
The functions $f_K(k_d)$ localise these modes in some region of spacetime and hence the mode operators  $\hat{b}_K$ can describe modes detected or generated by a local observer. Alice and Bob will integrate the photocurrent from their detectors over a time long compared to the inverse of the frequency being analysed (as will be determined by the frequency of their local oscillators). If the local oscillator amplitude satisfies $\beta >> 1$ then it can be treated as a classical field and the average value of the signal quadrature will be given by the expectation value:
\begin{eqnarray}
\langle X_{J, S}(\phi) \rangle &=& {{1}\over{\beta}} \langle \int d \tau \hat{O}_J(\tau, \phi) \rangle \nonumber \\
&=&  \int d \tau \langle 0| 
\int dk_d (e^{-i \phi} \hat b_{J, S}^{\dagger}  f_L(k_d, \tau) f_{J,D}^*(k_d, \tau_J)) \nonumber \\
&& \;\;\;\;\;\;\ + \; e^{i \phi} \hat b_{J, S} f^*_L(k_d, \tau) f_{J,D}(k_d, \tau_J)) 
|0\rangle 
\label{Xe}
\end{eqnarray}
where $f_L$ ($f_{J,D}$) are the detector (displacement) mode functions and $|0 \rangle$ is the Minkowski vacuum and $\tau_J$ refers to the conformal time around which the local oscillator pulse is centred where $J=F$ for Alice and $J=P$ for Bob. The detector mode functions are assumed identical for Alice and Bob. Eq.\ref{Xe} can be significantly simplified if we assume our detectors are broadband and are placed close to the focus of the local oscillator such that the paraxial approximation can be made. We obtain \cite{DOW13}
\begin{eqnarray}
\langle X_{J, S}(\phi) \rangle &\cong&  \langle 0| 
\int dk_{d1}
( e^{i \phi} f_D(k_{d1}) \hat b_{k_{d1}, J} \nonumber \\
&& \;\;\;\;\;\;\  + e^{-i \phi} f_D^*(k_{d1}) \hat b_{k_{d1}, J}^{\dagger}) |0\rangle
\label{Xs1}
\end{eqnarray}
where we have defined new boson annihilation operators:
\begin{equation}
\hat b_{k_{d1}, J} \equiv  \int d\vec k_{d} \;  g_S(\vec k_{d})\; \hat b_{k_d, J}
\end{equation}
where the transverse mode function of the detectors is given by $g_S(\vec k_{d}) = g_S(k_{d2}) h_S(k_{d3})$ (see the Appendix for more details).

In order to calculate the expectation value of Eq.\ref{Xs1} against the Minkowski vacuum we need to rewrite the measurement operators in terms of Minkowski modes. It was shown in Ref.\cite{OLS11} that the transformation relations between the Future and Minkowski spectral modes, appropriate for Alice, are identical to those between the right Rindler and Minkowski spectral modes, and can be written \cite{TAK86}
\begin{eqnarray}
\hat{b}_{F, k_d} &=& \int dk_s ( A_{k_dk_s}\delta(\vec{k}_d-\vec{k}_s) \hat{a}_{k_s}  \nonumber\\ 
&& \;\;\;\;\;\;+ B_{k_dk_s}\delta(\vec{k}_d+\vec{k}_s) \hat{a}_{k_s}^{\dagger})
\label{rob}
\end{eqnarray}
where the operators $\hat{a}_{k_s}$ are the plane wave Minkowski operators, obeying the usual boson commutation relation
\begin{equation}
[\hat{a}_{k_s}, \hat{a}_{k'_s}^{\dagger}] = \delta(k_{s1}-k'_{s1})\delta(k_{s2}-k'_{s2})\delta(k_{s3}-k'_{s3}),
\label{Mcomm}
\end{equation}
$\vec{k}_d=(k_{d2},k_{d3}), \vec{k}_s=(k_{s2},k_{s3})$, $\omega_s$ is the signal frequency $\omega_s=\sqrt{k_{s1}^2+k_{s2}^2+k_{s3}^2}$,
and the Bogolyubov coefficients are given by
\begin{eqnarray}
A_{k_dk_s} &=& \frac{1}{\sqrt{2\pi\omega_s}}e^{i \phi(k_{s1}) \Omega_{d}/a}  N(\Omega_{d}) \nonumber \\
B_{k_dk_s} &=& e^{-\pi \Omega_{d}/a} A_{k_dk_s}
\label{bcb}
\end{eqnarray}
where 
\begin{eqnarray}
|N(\Omega_{d})| &=& {{1}\over{\sqrt{1-e^{-2\pi \Omega_{d}/a}}}}
\end{eqnarray}
and
\begin{eqnarray}
\phi(k_{s1}) &=& {{1}\over{2}} \; ln \left(\frac{\omega_s+k_{s1}}{\omega_s-k_{s1}}\right).
\end{eqnarray}
For Bob the appropriate transformation is that between the Past and Minkowski modes which can be written
\begin{eqnarray}
\hat{b}_{P, k_d} &=& \int dk_s ( A^*_{k_dk_s}\delta(\vec{k}_d-\vec{k}_s) \hat{a}_{k_s}  \nonumber\\ 
&& \;\;\;\;\;\;+ B^*_{k_dk_s}\delta(\vec{k}_d+\vec{k}_s) \hat{a}_{k_s}^{\dagger})
\label{arob}
\end{eqnarray}
Given that the transformations of Eq.\ref{rob} and \ref{arob} are linear in the creation and annihilation operators and that, by definition, the Minkowski annihilation operator annihilates the Minkowski vacuum, i.e.
\begin{equation}
\hat{a}_{k_s} \;  |0 \rangle = 0, 
\label{ann}
\end{equation}
it follows that $\langle 0| \; \hat{b}_{k_{d1},S} \;  |0 \rangle = \langle 0| \; \hat{b}^{\dagger}_{k_{d1},S} \;  |0 \rangle = 0$, and hence $\langle X_{R, S}(\phi) \rangle = \langle X_{A, S}(\phi) \rangle = 0$, i.e. the average values of the homodyne signals are zero.

More interesting is to calculate the variances and correlation variances of the signals. The variances of the individual signals (given that the first order expectation values are zero) are given by
\begin{eqnarray}
V_J(\phi) &=& {{1}\over{\beta^2}} \langle( \int d \tau \hat{O}_J(\tau, \phi))^2 \rangle. 
\label{Vs2}
\end{eqnarray}
We now transform to Minkowski modes. For Alice, using Eq.\ref{rob} we find
\begin{eqnarray}
\hat b_{k_{d1},F} &=& \int d\vec k_{d} \;  g_S(\vec k_{d})\int dk_s( A_{k_dk_s} \hat{a}_{k_s}\delta(\vec{k}_d-\vec{k}_s) \nonumber \\
& + & B_{k_dk_s} \hat{a}_{k_s}^{\dagger}\delta(\vec{k}_d+\vec{k}_s)) \nonumber \\
&=& \int dk_{s} \frac{ 1}{\sqrt{2\pi\omega_s(e^{2\pi \bar \Omega_{d}/a}-1)}}\left(\frac{\omega_s+k_{s1}}{\omega_s-k_{s1}}\right)^{i\frac{1}{2} \bar \Omega_{d}/a}
\nonumber \\
&& \;\;\; \times (e^{\pi \bar \Omega_{d}/a} \hat{a}_{k_s}g_S(\vec k_{s}) + \hat{a}_{k_s}^{\dagger}g_S(-\vec k_{s}))
\label{F}
\end{eqnarray}
where $\bar \Omega_d = \sqrt{k_{d1}^2 +\vec k_s^2}$. 
Similarly for Bob (Eq.\ref{arob}) we find
\begin{eqnarray}
\hat b_{k_{d1},P} &=& \int d\vec k_{d} \;  g_S(\vec k_{d})\int dk_s( A^*_{k_dk_s} \hat{a}_{k_s}\delta(\vec{k}_d-\vec{k}_s) \nonumber \\
& + & B^*_{k_dk_s} \hat{a}_{k_s}^{\dagger}\delta(\vec{k}_d+\vec{k}_s)) \nonumber \\
&=& \int dk_{s} \frac{ 1}{\sqrt{2\pi\omega_s(e^{2\pi \bar \Omega_{d}/a}-1)}}\left(\frac{\omega_s+k_{s1}}{\omega_s-k_{s1}}\right)^{-i\frac{1}{2} \bar \Omega_{d}/a}
\nonumber \\
&& \;\;\; \times (e^{\pi \bar \Omega_{d}/a} \hat{a}_{k_s}g_S(\vec k_{s}) + \hat{a}_{k_s}^{\dagger}g_S(-\vec k_{s}))
\label{P}
\end{eqnarray}
Substituting this into Eq.\ref{Vs2}, using the properties of the Minkowski modes, Eq.\ref{Mcomm} and Eq.\ref{ann}, the identity
\begin{eqnarray}
 \int dk_{s1} \frac{ 1}{2\pi\omega_s}\left(\frac{\omega_s+k_{s1}}{\omega_s-k_{s1}}\right)^{i\frac{1}{2}(x-x')} = \delta(x-x')
\end{eqnarray}
and making a change of variables to $\bar \Omega_d$, we find for both Alice and Bob
\begin{eqnarray}
V_J(\phi) &=&  \int_{-\infty}^\infty d\vec k_s \int_{|\vec k_s|}^\infty d\bar \Omega_{d} \times \nonumber \\
&&  |\bar f_D(\bar \Omega_{d},\vec k_{s})g_{J,S}(\vec k_{s})|^2 \frac{e^{2\pi \bar \Omega_{d}/a}+1}{e^{2\pi \bar \Omega_{d}/a}-1} 
\label{Vs3}
\end{eqnarray}
where
\begin{eqnarray}
\bar f_{D}(\bar \Omega_{d},\vec k_{s}) = {{\sqrt{K}\bar \Omega_d}\over{\sqrt{\bar \Omega_d^2 - \vec k_{s}^2}}}(f_{D}(|k_{d1}|) + f_{D}(-|k_{d1})|)
\label{el1}
\end{eqnarray}
and $K$ is a normalization constant satisfying
\begin{eqnarray}
\int_{-\infty}^\infty d\vec k_s \int_{|\vec k_s|}^\infty d\bar \Omega_{d}  |\bar f_D(\bar \Omega_{d},\vec k_{s})g_{S}(\vec k_{s})|^2= 1 .
\label{N}
\end{eqnarray}

We can obtain an approximate solution to Eq.\ref{Vs3} by assuming that Alice's or Bob's longitudinal local oscillator mode function, 
\begin{eqnarray}
f_{J,D}(k_{d1}) = \sqrt{{{1}\over{\sqrt{d \pi}}}} e^{-{{(k_{d1}-k_{do})^2}\over{2 d}}} e^{i \epsilon_J k_{d1}}
\label{del1}
\end{eqnarray}
is a Gaussian, strongly peaked around the frequency $k_{do}$ with respect to their conformal time, $\tau_J$, where we have assumed $|\epsilon_J| <<1$, and that its transverse function, 
\begin{eqnarray}
g_{J,S}(\vec k_{s}) = \sqrt{{{1}\over{\sqrt{s \pi}}}} e^{-{{(\vec k_{s})^2}\over{2 s}}} 
\label{del2}
\end{eqnarray}
is also a Gaussian, strongly peaked around zero frequency, such that we can make the approximations $|f_{J,D}(k_{d1})|^2 \approx \delta(k_{d1} - k_{do})$ and $|g_{J,S(}\vec k_{s})|^2 \approx \delta(\vec k_{s})$
and hence obtain the approximate expression for the variance of the signal mode
\begin{eqnarray}
V_J \approx  \frac{ e^{2\pi \Omega_{do}/a}+1}{e^{2\pi \Omega_{do}/a}-1}
\label{Vs4}
\end{eqnarray}
where $\Omega_{do}= |k_{do}|$. This expression is identical to that obtained for homodyne detection, by an inertial observer, of a thermal bath at temperature $T=a\hbar/(2 \pi k)$, with $k$ Boltzmann's constant, as expected from the Unruh effect. 
\begin{figure}[htb]
\begin{center}
\includegraphics*[width=8cm]{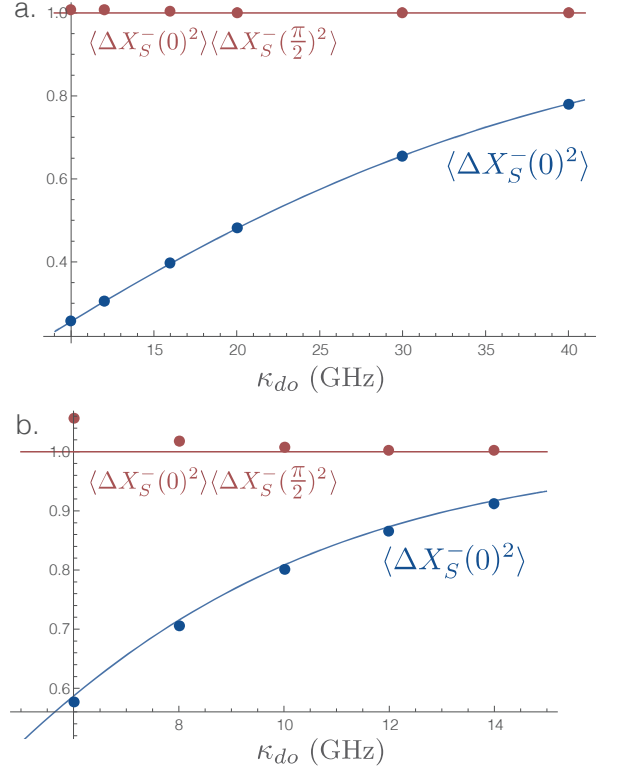}
\caption{Squeezing correlation (blue line) and uncertainty product (red line) predicted for observation of the space-time vacuum with energy-scaled homodyne detection. The system is entangled if the squeezing correlation is less than $1$. The system is pure if the uncertainty product is equal to $1$. The blue line (red line) is the approximate solution whilst the dots are corresponding numerical solutions of the exact expressions.  The parameters are: (a) $s=0.25 \times 10^9 s^{-1}$, $d=2.0 \times 10^9 s^{-1}$, $a=60 \times 10^9 s^{-1}$; and (b) $s=0.5 \times 10^9 s^{-1}$, $d=5.0 \times 10^9 s^{-1}$, $a=14 \times 10^9 s^{-1}$.
}
\label{fig1}
\end{center}
\end{figure}

The correlation and anti-correlation variances of the signal modes between Alice and Bob are given by
\begin{eqnarray}
\langle \Delta X_{S}^{\pm}(\phi)^2 \rangle &=& \langle (\int d \tau (\hat{O}_F(\tau, \phi)\pm \hat{O}_P(\tau, \phi))^2 \rangle/(2\beta^2) \nonumber \\
&=& V_F + V_P \nonumber \\
&\pm& 2 \langle (\int d \tau (\hat{O}_F(\tau, \phi) \hat{O}_P(\tau, \phi)) \rangle/(2\beta^2) \nonumber \\
\label{VC}
\end{eqnarray}
The variance in the correlations gives us a sufficient condition for entanglement in the field. In particular if 
\begin{eqnarray}
\langle \Delta X_{S}^{+}(\phi)^2 \rangle \langle \Delta X_{S}^{-}(\phi+{{\pi}\over{2}})^2 \rangle < 1
\label{Ent}
\end{eqnarray}
for some choice of $\phi$ then the joint state observed by Rob and Alice was entangled. Furthermore if in addition we have
\begin{eqnarray}
&& \langle \Delta X_{S}^{-}(\phi)^2 \rangle \langle \Delta X_{S}^{-}(\phi+{{\pi}\over{2}})^2 \rangle  \nonumber \\
&=& \langle \Delta X_{S}^{+}(\phi)^2 \rangle \langle \Delta X_{S}^{+}(\phi+{{\pi}\over{2}})^2 \rangle = 1
\label{pure}
\end{eqnarray}
then the state detected was pure. Evaluating the cross-correlation term for $\phi=0$  
gives:
\begin{eqnarray}
&& \langle (\int d \tau (\hat{O}_F(\tau, 0) \hat{O}_P(\tau, 0)) \rangle/(2\beta^2) \nonumber \\
&=&   \int_{-\infty}^\infty d\vec k_s \int_{|\vec k_s|}^\infty d\bar \Omega_{d}  \{\bar f_{D,F}(\bar \Omega_{d}, \vec k_s)\bar f_{D,P}(\bar \Omega_{d}, \vec k_s)  \nonumber \\
&\times& g_{S,F}(\vec k_{s})g_{S,P}(-\vec k_{s}) e^{-i(\tau_F - \tau_P)} \nonumber \\
& + & \bar f^*_{D,F}(\bar \Omega_{d}, \vec k_s)\bar f^*_{D,P}(\bar \Omega_{d}, \vec k_s)  \nonumber \\
&\times& g_{S,F}^*(-\vec k_{s})g_{S,P}^*(\vec k_{s})\} e^{i(\tau_F - \tau_P)}  \frac{e^{\pi \bar \Omega_{d}/a}}{e^{2\pi \bar \Omega_{d}/a}-1} 
\label{cross1}
\end{eqnarray}
and for $\phi = \pi/2$
\begin{eqnarray}
&& \langle (\int d \tau (\hat{O}_F(\tau, \pi/2) \hat{O}_P(\tau, \pi/2)) \rangle/(2\beta^2) \nonumber \\
&=&  - \int_{-\infty}^\infty d\vec k_s \int_{|\vec k_s|}^\infty d\bar \Omega_{d}  \{\bar f_{D,F}(\bar \Omega_{d}, \vec k_s)\bar f_{D,P}(\bar \Omega_{d}, \vec k_s)  \nonumber \\
&\times& g_{S,F}(\vec k_{s})g_{S,P}(-\vec k_{s}) e^{-i(\tau_F - \tau_P)} \nonumber \\
&+& \bar f^*_{D,F}(\bar \Omega_{d}, \vec k_s)\bar f^*_{D,P}(\bar \Omega_{d}, \vec k_s)  \nonumber \\
&\times& g_{S,F}^*(-\vec k_{s})g_{S,P}^*(\vec k_{s})\} e^{i(\tau_F - \tau_P)}  \frac{e^{\pi \bar \Omega_{d}/a}}{e^{2\pi \bar \Omega_{d}/a}-1}  
\label{cross2}
\end{eqnarray}
To maximize the observed entanglement we should impose a number of anti-symmetry conditions on the local oscillator mode functions of Alice and Bob. In particular we require 
$ \bar f_{D,F}(\bar \Omega_{d}, \vec k_s) = \bar f^*_{D,P}(\bar \Omega_{d}, \vec k_s)$ and 
$g_{F,S}^*(-\vec k_{s})=g_{P,S}(\vec k_{s})$. Notice that the mode functions introduced in Eqs \ref{del1}, and \ref{del2} satisfy these conditions provided 
$\epsilon_F = -\epsilon_P$. With this condition, plus having $\tau_F = -\tau_P$ and assuming as in the previous section that Alice and Bob's local oscillator mode functions are strongly peaked we obtain
\begin{eqnarray}
\langle \Delta X^-_S(0)^2 \rangle = \langle \Delta X^+_S({{\pi}\over{2}})^2 \rangle \approx \frac{ e^{\pi \Omega_{do}/a}-1}{e^{\pi \Omega_{do}/a}+1}
\label{Vs4}
\end{eqnarray}
thus fulfilling the condition of Eq.\ref{Ent} that the detected fields are entangled for $\phi = 0$ and all $a>0$. In addition it is straightforward to show that 
\begin{eqnarray}
 \langle \Delta X_{S}^{-}(0)^2 \rangle \langle \Delta X_{S}^{-}({{\pi}\over{2}})^2 \rangle
= \langle \Delta X_{S}^{+}(0)^2 \rangle \langle \Delta X_{S}^{+}({{\pi}\over{2}})^2 \rangle \approx 1  \nonumber \\
\label{pure2}
\end{eqnarray}
In Fig.1 we plot the approximate solutions Eqs \ref{Vs4} and \ref{pure} and numerical evaluations of Eq.\ref{VC} using Eqs \ref{cross1} and \ref{cross2} -- this shows that there are parameters for which the approximate solutions agree well with the exact ones. Thus under the described conditions it is in principle possible to observe very pure entanglement from the quantum vacuum via scaled homodyne detection. We have so far developed this theory in conformal time co-ordinates. After explaining our QKD protocol we will investigate the relationship between conformal and laboratory parameters, making explicit the connection to laboratory time and discussing the feasibility of the necessary parameters.
\begin{figure}[htb]
\begin{center}
\includegraphics*[width=8cm]{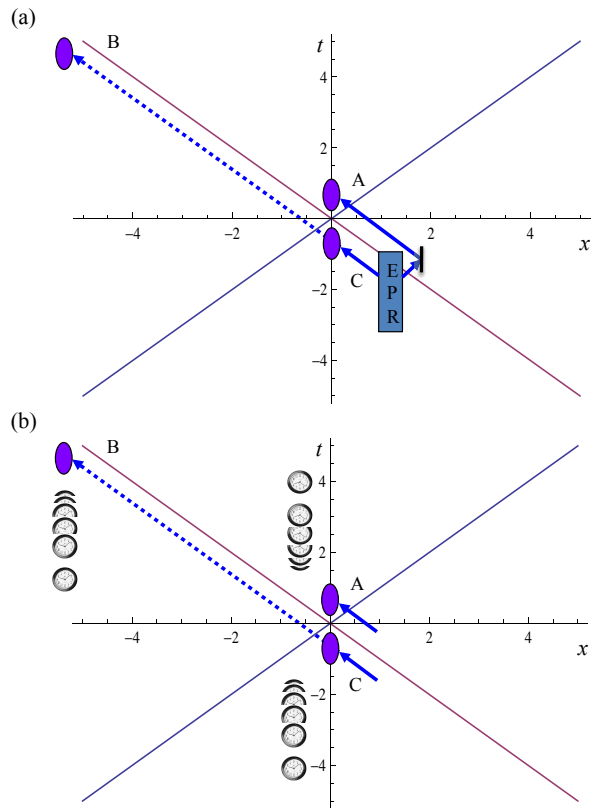}
\caption{Space time diagram representation of (a) standard continuous variable quantum key distribution and (b) quantum key distribution without sending a quantum signal. The points $A$ and $C$ lie in Alice's station whilst point $B$ lies in Bob's station. In both cases no measurement is made at point $C$ when the QKD protocol is being enacted. The unequally spaced clocks in (b) represent the energy scaled homodyne detection.
}
\label{fig2}
\end{center}
\end{figure}
\vspace{.3cm}

\section{QKD without sending a quantum signal} 
We now wish to show that, using the measurement techniques described above, it is possible to exchange a secret quantum key \cite{Scarani:2009p378} without sending a quantum signal. It is well known that entanglement based versions of QKD are possible \cite{ek91}. We are particularly interested in the entanglement based version of continuous variable QKD \cite{Grosshans:2003p526}. The important point is that entanglement based continuous variable QKD relies on the distribution of a two-mode squeezed vacuum state. However, the correlations observed by the Future/Past detectors of Alice and Bob above are exactly of the form of a two-mode squeezed vacuum state. The question then is whether it possible to observe these correlations at a large spatial separation as required for a QKD protocol. In the following we answer this question in the affirmative.

Consider the entanglement based continuous variable QKD protocol depicted in Fig.2(a). A two-mode squeezed state is produced by the Einstein Podolsky Rosen (EPR) source as a pair of counter-propagating pulses. By reflecting one of the pulses they can be arranged to arrive at the spatial origin at ~ $\pm t $ (points $A$ and $C$).  Assuming the EPR source is lossless, the correlations between an ensemble of such pulses will be given by \cite{BAC03}
\begin{eqnarray}
\langle \Delta X^-_S(0)^2 \rangle = \langle \Delta X^+_S({{\pi}\over{2}})^2 \rangle = (\sqrt{G} - \sqrt{G-1})^2
\label{EPR}
\end{eqnarray}
where $G \ge 1$ is the gain of the EPR source. The correlations also satisfy Eq.\ref{pure}. The source is thus entangled and pure for $G>1$. In order to perform QKD Alice only detects the EPR state at point $A$ and allows the other beam to propagate to Bob who detects it at point $B$. As we move along the geodesic from Alice to Bob the transverse mode will expand. Assuming a Gaussian mode with its focus at $C$, then at $B$, a long distance from $C$, the wave fronts will be approximately spherical and centred on $C$. However, if we assume Bob's detector has the same radius as Alice's then the wave-front curvature across Bob's detector is negligible and the paraxial approximation can still be made. This means the longitudinal part of the wave function is unchanged. On the other hand the intensity of the field on axis has dropped by a factor of $(z_o/z)^2$, where $z_o$ is the Rayleigh length and $z$ is the distance between the points $C$ and $B$ \cite{BAC03}. Putting this all together one concludes that Bob will detect a field at $B$ with the same longitudinal spectral structure as that at $C$ but with attenuated correlations given by
\begin{eqnarray}
\langle \Delta X^-_S(0)^2 \rangle = \langle \Delta X^+_S({{\pi}\over{2}})^2 \rangle = \eta(\sqrt{G} - \sqrt{G-1})^2 + 1 - \eta \nonumber \\
\label{EPRa}
\end{eqnarray}
where $\eta = (z_o/z)^2$. It is well known that such correlations between measurements made by Alice and Bob are sufficient to allow them to implement QKD \cite{Grosshans:2003p526, WEE11}. The protocol is: (i) Alice generates a large ensemble of pairs of pulses from an EPR source, sending one of each pair to Bob; (ii) Alice and Bob randomly, and independently choose to measure either $X^-_S(0)$ or $X^-_S(\pi/2)$ on each of the pulses; (iii) they sift their results to identify occasions upon which they both measured the same quadrature angles; (iv) they release a subset of their data in order to characterize the channel (between $C$ and $B$); and (v) if the channel is of sufficiently high quality such as to give them an information advantage over any possible eavesdropper they proceed to extract a secret key from the remaining data using a reconcilliation procedure. In principle such a protocol, with correlations of the form of Eq.\ref{EPRa}, allows secret key to be distributed over arbitrarily long distances \cite{Grosshans:2003p2402}. Under realistic conditions, $\eta$'s as low as a few percent still allow key to be distributed \cite{jouguet13}.

Now consider Fig.2(b) depicting the detection of vacuum entanglement at positions $A$ and $C$ as described in the previous section. A comparison of Eq.\ref{Vs4} and \ref{EPR} shows that the correlations observed from the EPR source are the same as those predicted for conformal time detection of the vacuum provided 
\begin{eqnarray}
G=e^{2\pi \Omega_{do}/a}/(e^{2\pi \Omega_{do}/a}-1).
\label{EPRvac}
\end{eqnarray}
Now consider moving the second detector to point $B$. We can calculate the new correlations between $A$ and $B$ by translating the coordinates in Eq.\ref{P} along the geodesic from $C$ to $B$. Given that Eq.\ref{P} is expressed in terms of Minkowski modes, the translation behaves in the same way as for the EPR source, i.e. preserving the longitudinal mode function along the geodesic but expanding transversally \cite{note}. Hence the correlations between points $A$ and $B$ are given by Eq.\ref{EPRa} with $G$ given by Eq.\ref{EPRvac}. Thus Alice and Bob can observe the same correlations from the vacuum entanglement at points $A$ and $B$ as they would have from distributing an EPR state between these points. As a result they can implement a QKD protocol based on those correlations without having to send a quantum signal between them. In particular the new protocol is: (i) Alice and Bob agree on a sequence of time intervals in which they will observe the vacuum at their local positions using their conformal detectors tuned to maximize the observed correlations; (ii) Alice and Bob randomly, and independently choose to measure either $X^-_S(0)$ or $X^-_S(\pi/2)$ in each time window; the rest of the protocol then follows (iii) - (v) above.

We can now see why the directionality of the detection is key to observing the vacuum entanglement at large separations. According to Ref. \cite{OLS11}, the entanglement at points $A$ and $C$ in Fig.2(b) is isotropic, i.e. it is maximized for both left and right propagating fields. However, when we move along the null-geodesic to point $B$, this symmetry is broken -- now only the left propagating fields are entangled (the corresponding point of maximal entanglement for the right movers is the mirror image of point $B$ on the right-side of Fig.2(b)). Without the directionality of the detection the entanglement  would rapidly disappear as we moved along the null-geodesic.

\section{More realistic parameters} 

We have shown that in principle QKD can be carried out without exchanging a quantum signal by exploiting the intrinsic entanglement of the space-time vacuum. This result relies on the equivalence between the approximate solution of Eq.\ref{Vs4} and the exact numerical solution of Eq.\ref{VC} as demonstrated in Fig.1(a). However, the value of $a$ used to obtain Fig.1(a) is impractically large.  We now explore the effect of using a more realistic scaling parameter in the protocol. In particular we would like the required energy scaling not to exceed approximately an order of magnitude and the rate of change not to be too extreme. A key relationship is that between time intervals in the lab frame, $\Delta t$ and those in conformal time, $\Delta \tau$. Consider first Alice's future mode. We have:
\begin{eqnarray}
\Delta t = {{e^{a \tau_o}}\over{a}}(e^{a \Delta \tau} - 1)
\label{t}
\end{eqnarray}
where $\tau_o$ is the initial conformal time. The total time interval over which the conformal detectors should integrate is determined by the temporal width of the pulse in conformal time: $\Delta \tau_T > 1/\sqrt{d}$. This in turn determines the ratio between the initial and final frequencies in the lab frame: 
\begin{eqnarray}
{{\omega_f}\over{\omega_i}} = e^{-a \Delta \tau_T}
\label{w}
\end{eqnarray}
Another useful relationship is that between the conformal frequency and the initial lab frequency:
\begin{eqnarray}
{{\Omega_{do}}\over{\omega_i}} = e^{a \tau_o}
\label{Ow}
\end{eqnarray}
which leads to a time dependent frequency given by
\begin{eqnarray}
\omega(\Delta t) = \frac{\Omega_{do}}{e^{a\tau_o} + a\Delta t}.
\label{t-m}
\end{eqnarray}
Here $\Delta t = t - t_i$ where the initial lab time is $t_i = a^{-1} e^{a \tau_o}$. Eq. \ref{t-m} can also describe the time dependency of Bob's past mode detector but now $\Delta t = t _f- t$ where the final lab time is $t_f = a^{-1} e^{a \tau_o}$. Notice that $t=0$ in the lab frame correspond to $\tau \to -\infty$ in conformal time. Hence substituting $\tau_o \to -\infty$ in Eq.\ref{t-m} recovers the $1/a t$ scaling of the Hamiltonian in the Schr\"odinger Equation (Eq.\ref{SE}).

The strength of the vacuum entanglement depends on $\Omega_{do}/a$ (see Eq.\ref{EPRvac}) with values less than $1/(2 \pi)$ leading to non-negligible entanglement. Putting these conditions together leads to various compromises in order to simultaneously obtain strong entanglement with reasonable parameters. One possibility is the choice: $a = 14 \times 10^{9} s^{-1}$ (parameterisation of the difference between Minkowski and proper time); $d = 5 \times 10^{9} s^{-1}$ (width in frequency of the detectors); and $s = 0.5 \times 10^{9} s^{-1}$ (width in frequency of the signal). Table I shows the resulting lab frame parameters for detector strongly peaked around a frequency $\Omega_{do} = 10 \times 10^{9}  \; rad \; s^{-1}$ and various different initial times $\tau_o$. The final column labeled $T_{max}$ estimates the highest allowable background temperature such that, given a particular $\omega_f$, it is still a good approximation to take the surrounding space-time to be in the vacuum state.

The behaviour of the correlations as a function of conformal frequency for this choice of parameters is shown in Fig.1(b). Observe that there are now regions in which the observed correlations are no longer pure. This occurs because the restrictions in choosing parameters leads to some smearing in frequency of the correlations. The reduction in purity should be attributed to a possible eavesdropper and so must be taken into account when calculating the secret key rates by attributing additional imperfections to the channel. We calculate the key rate in the long key-length limit against arbitrary attacks \cite{Navascues:2006p805,GarciaPatron:2006p381,Renner:2009p1}. From the effective covariance matrix defined by the observed correlations, and using the straightforward secret key formulas provided in Ref.\cite{WEE11}, plots of secret key rates for conformal frequency $\Omega_{do} = 40 \times 10^{9} \; rad \; s^{-1}$ ($\Omega_{do} = 10 \times 10^{9}  \; rad \; s^{-1}$) and $a = 60 \times 10^{9} s^{-1}$ ($a = 14 \times 10^{9} s^{-1}$) are given in Fig.3. The plots show that secret key can still be distributed over significant distances using the more realistic parameters of Fig.1(b) and Table I.
\begin{table}
\begin{tabular}{|c|c|c|c|c|}
\hline
$\tau_o$ & $\omega_i$ & $\omega_f$ & $\Delta t $ & $T_{max}$  \\
\hline\hline
$-0.98 \; ns$ & $9.40 \times 10^{15}  \; rad \; s^{-1}$ &  $6.28 \times 10^{14}  \; rad \; s^{-1}$ & $10 \; fs$ & $300K$ \\
\hline
$-0.47 \; ns$ & $7.48 \times 10^{12}  \; rad \; s^{-1}$ & $5.0 \times 10^{11}  \; rad \; s^{-1}$ & $12.6 \; ps$ & $3K$  \\
\hline
$-0.14 \; ns$ & $7.48 \times 10^{10}  \; rad \; s^{-1}$ & $5.0 \times 10^{9}  \; rad \; s^{-1}$ & $1.26 \; ns$ & $10^{-3} K$  \\
\hline
\end{tabular}
\caption{Values of the lab frame parameters given particular choices of $\tau_o$ with $a = 14 \times 10^{9} s^{-1}$ and $\Omega_{do} = 10 \times 10^{9}  \; rad \; s^{-1}$. $T_{max}$ is an approximate indication of the maximum background temperature consistent with the chosen parameters.}
\label{exptable}
\end{table}
\begin{figure}[htb]
\begin{center}
\includegraphics*[width=8cm]{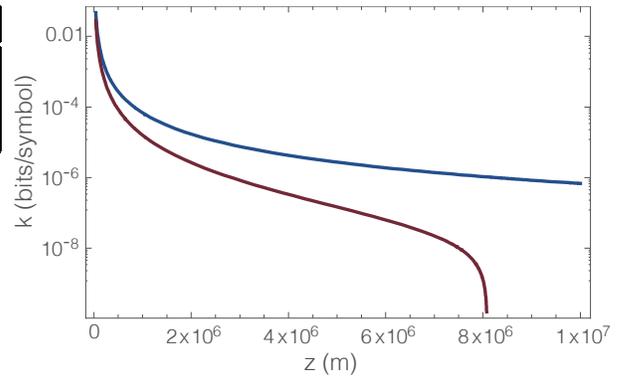}
\caption{Secret key rate normalized to the raw key rate, $K$, plotted against distance between Alice and Bob in metres. The value of the Rayleigh length is calculated using $z_0 = \pi W^2/\lambda$ where we assume a waist size of $W = 19.25 cm$ and a wavelength of $\lambda = 3 \mu m$ (corresponding to $\omega_f = 6.28 \times 10^{14}  \; rad \; s^{-1}$ \cite{note}). The two graphs correspond to different values of $a$. The blue line is for $a=60 \times 10^9 s^{-1}$ (as per Fig.1(a)), whilst the red line is for the more realistic value of $a=14 \times 10^9 s^{-1}$ (as per Fig.1(b) and Table I). The plot shows that even with the more realistic value for $a$ absolutely secure key can be shared over distances of thousands of kilometres.
}
\label{fig3}
\end{center}
\end{figure}

\section{Discussion} 

We have shown that by employing energy-scaled homodyne detection it is possible for two observers, separated by non-trivial distances, to measure the correlations of entangled space-time vacuum modes with sufficient quality so as to be able to implement a QKD protocol. The energy levels of the detectors must change with lab time according to Hamiltonians obeying Eq.\ref{SE}. The local oscillators should also follow this scaling. The more realistic parameters identified in the previous section are still extremely challenging, requiring the energy levels to scale over an order of magnitude in frequency. At room temperature we are required to work at optical to near infra-red frequencies and the scaling must happen over femto-second time scales (top line of Table 1). In space we might work in the far infra-red, with scaling on pico-second time-scales (middle line of Table 1). Though not practical for communications, perhaps the most realistic near term prospects for demonstrating the basic principles of this effect would be to work at microwave frequencies in a dilution refrigerator (bottom line of Table 1). Strictly, the theory presented here only applies in vacuum. The effects of birefringence or atmospheric absorption lines would need to be included for earthbound communication. Various methods might be employed to induce the required detector energy scaling - see the Appendix for a brief discussion of some possibilities.

The techniques discussed here are expected to generalize to various other quantum communication protocols that rely on the distribution of EPR type entanglement such as continuous variable teleportation and continuous variable dense coding -- also enabling such protocols to be carried out in a relativistic scenario without requiring the sending of any, or as many, quantum signals. 

\section{Appendix} 

{\bf Intensity detection:} The homodyne detector is constructed from two macroscopic intensity detectors. It is well known that the excitation probability of an Unruh-DeWitt detector with resonant frequency $\omega$, weakly coupled to a scalar field, is proportional to the expectation value of the Hermitian operator $\hat a_{k}^{\dagger} \hat a_k$, where $\hat a_k$ is a single frequency ($k=|\omega|$) Minkowski annihilation operator for the field and the expectation value is taken over the state of the field. In quantum optics the measurement operator for a macroscopic intensity detector can be accurately modelled by Hermitian operators constructed from coherent superpositions of $\hat a_k$ of different frequencies and their conjugates. For example, a broad-band intensity detector with a sharp time resolution can be modelled by the operator $\hat a^{\dagger}(t) \hat a(t)$ where $\hat a(t) = {{1}\over{\sqrt{2 \pi}}} \int dk e^{-i \omega t} \hat a_k$. This model will be accurate provided the envelope of the wave-packet of the detected pulse is slowly varying compared to the temporal resolution of the detector. Typically the the photo-current of the detector will be integrated over the pulse length giving the measurement operator $\int dt \; \hat a^{\dagger}(t) \hat a(t)$.

In Ref.\cite{OLS11} and \cite{OLS12} it was shown that the excitation probability of an Unruh-DeWitt detector with a time dependent resonant frequency corresponding to the Schr\"odinger equation Eq.\ref{SE} is proportional to the expectation value of $\hat b_{k_d}^{\dagger} \hat b_{k_d}$, where $\hat b_{k_d}$ is a single frequency Rindler annihilation operator whose relationship to the Minkowski operators is given by Eqs.\ref{rob} (for future modes) and \ref{arob} (for past modes). Thus a macroscopic, broad-band intensity detector whose constituent atomic absorbers have resonant frequencies which follow Eq.\ref{SE} will be well approximated by the measurement operator $\hat b^{\dagger}(\tau) \hat b(\tau)$ where 
\begin{equation}
\hat b(\tau) = {{1}\over{\sqrt{2 \pi}}} \int dk_d e^{-i \Omega_d \tau} \hat b_{k_d}.
\label{sl}
\end{equation}
%

\begin{figure}[htbp]
\centering
\includegraphics[width = 6cm]{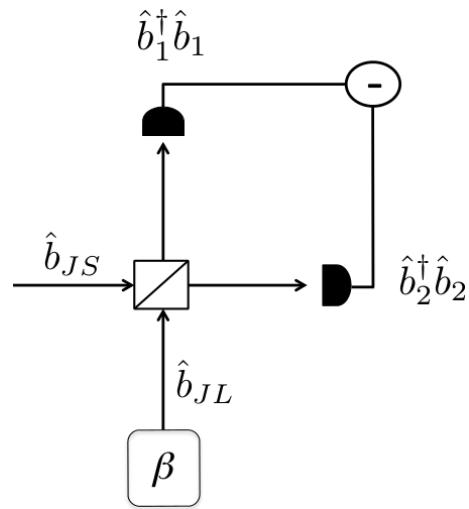}
\caption{\label{homoschem} Schematic of a homodyne detection to make a quadrature measurement. The signal mode $\hat{b}_{J,S}$ is combined with the local oscillator mode $\hat{b}_{\mathrm{J,L}}$ and then an intensity measurement is performed. Taking the difference of the two measurements gives an output that is directly proportional to the desired quadrature.}
\end{figure}

{\bf Homodyne Detection:} The homodyne detection is modelled by the standard setup shown in Figure \ref{homoschem}, except that the intensity detectors are the energy scaled ones described in the previous paragraphs and the displacement is by a chirped coherent field whose centre frequency scales in the same way as the detectors. The photo-currents from the two intensity detectors are subtracted to give a photo-current represented by the operator:  
\begin{eqnarray}
\hat b_1^{\dagger}(\tau) \hat b_1(\tau) -  \hat b_2^{\dagger}(\tau) \hat b_2(\tau).
\label{M2}
\end{eqnarray}
The signal mode $\hat{b}_{J,S}$ and local oscillator mode $b_{J,L}$ are interfered on a balanced beamsplitter before detection. The mixing on the beamsplitter is modelled by the operator transformations:
\begin{eqnarray}
\hat{b}_1 = \frac{\hat{b}_{J,S} + \hat{b}_{J, L}}{\sqrt{2}},  \hspace{5mm} \hat{b}_2 = \frac{\hat{b}_{J,S} - \hat{b}_{J,L}}{\sqrt{2}}.
\label{M3}
\end{eqnarray}
Substituting the transformations of Eq.\ref{M3} into Eq.\ref{M2} and integrating over time gives Eq.\ref{OR}.

The local oscillator mode is displaced by an amplitude $\beta$ relative to a mode characterized by the function $f_D(k_d, \tau)$. Given this classical characterization of the mode and amplitude the displacement can be carried out via local manipulations \cite{BAC03}. The operator transformation representing the displacement is:
\begin{equation}
\hat D_J^{ \dagger}(\beta) \hat b_{J,L} \hat D_J(\beta) = \hat b_{J,L} + \beta  \int dk_d f_{J,L}(k_d, \tau)   f_D^*(k_d, \tau). 
\label{D}
\end{equation}
Making this transformation to Eq.\ref{OR} and using the assumption that $\beta>>1$ to neglect terms not multiplied by $\beta$ leads to Eq.\ref{Xe}. Notice that all terms depending on quantum features of the local oscillator mode are neglected at this point, showing that the local oscillator only plays the role of a classical phase reference. Also notice that we express the mode function of the local oscillator in terms of Future or Past co-ordinates, anticipating that we will ultimately choose this function to represent an approximate single frequency mode in these co-ordinates (and hence a chirping pulse in the lab frame).
Finally we choose the detector mode functions to be given according to Eq.\ref{sl}. Integrating over time then leads to Eq.\ref{Xs1}. 

{\bf Scanning Frequencies:} The scheme requires the resonant frequencies of the absorbers forming the detectors to change rapidly in time according to the recipe of Eq. \ref{t-m}. At optical frequencies, atomic resonances can be scanned by the application of an electric field through the Stark effect, though the changes in frequency usually induced are far smaller than required here. Artificial absorbers, such as quantum dots, can also be scanned via the application of an electric field, but again probably not over the ranges required here. The resonant frequencies of quantum dots can be changed over broader frequency ranges by modifying their physical characteristics, though how this could be achieved rapidly is unclear. Perhaps one possibility might be to engineer a non-homogeneous detector in which there was a spatial gradient in the resonant frequencies. By rapidly sweeping the incoming pulse over the spatial gradient an effective, time varying detection frequency might be achieved. 

The creation of the local oscillators at fixed conformal frequencies can be described as follows:
\newline
1) Alice and Bob make independent unitary displacements of the Minkowski vacuum so as to each make phase-locked (Minkowksi) coherent states with amplitudes $\beta$.
\newline
2) They independently chirp their local oscillators. Alice introduces a chirp with a descending frequency in time that approximately corresponds to a single frequency in Future conformal time. Bob introduces a chirp with a rising frequency in time that approximately corresponds to a single frequency in Past conformal time. The specific dependence of frequency on elapsed lab time is given by Eq. \ref{t-m}.
\newline
This chirping of the local oscillator pulses would also be challenging to achieve as again standard techniques such as electro-optic or acousto-optic modulation would probably not cover sufficient frequency range at optical frequencies. 

At micro-wave frequencies the technical challenges are perhaps less daunting with broader tuning ranges available with artificial absorbers such as super-conducting qubits \cite{SEM10, SEM11} and the slower rate at which the scanning must be carried out (see Table I). Broad and rapid frequency tuning of coherent sources of micro-waves are also available.


\begin{thebibliography}{99}
\bibitem{note} A minor complication is that the Rayleigh length is inversely proportional to the wavelength. As a result, assuming that the local oscillator is designed in such a way that the waist of its mode has a constant size for Alice, the spot size observed by Bob at a distant point will expand as the local oscillator frequency is scanned. When analyzing the secret key rate we assume the worst case scenario that the attenuation rate is given by the Rayleigh length calculated from the final frequency $\omega_f$. This strategy thus uses an upper bound on the attenuation rate of the channel. The average attenuation rate will always be less than this amount.
\end{thebibliography}
\end{document}